\documentclass[aps,prl,twocolumn,eqsecnum,superscriptaddress,floatfix,10pt]{revtex4-2}
\usepackage{amsmath,amssymb,amsfonts,stmaryrd,wasysym,graphicx,multirow,color,textcomp,subfigure,nicefrac,float,enumitem}
\usepackage{url}
\usepackage[colorlinks=true,citecolor=blue,urlcolor=blue,linkcolor=blue] {hyperref}

\usepackage{amsmath}
\usepackage{braket}
\usepackage{romannum}

\usepackage{color,soul} 
\usepackage{bm}
\usepackage[normalem]{ulem}

\def\be{\begin{equation}}
\def\ee{\end{equation}}
\def\bea{\begin{eqnarray}}
\def\eea{\end{eqnarray}}

\newcommand{\cyan}[1]{{\color{cyan}{\it {#1}}}} 

\counterwithout{equation}{section}

\begin{document}

\title{Stacking-dependent Topological Electronic Structures \\in Honeycomb-kagome Heterolayers}

\author{Chan Bin Bark}
\address{Department of Physics, Hanyang University, Seoul 04763, Republic of Korea}

\author{Hanbyul Kim}
\address{Department of Physics, Hanyang University, Seoul 04763, Republic of Korea}

\author{Seik Pak}
\address{Department of Physics, Hanyang University, Seoul 04763, Republic of Korea}

\author{Hong-Guk Min}
\affiliation{Department of Physics, Sungkyunkwan University, Suwon 16419, Korea}

\author{Sungkyun Ahn}
\affiliation{Department of Physics, Sungkyunkwan University, Suwon 16419, Korea}

\author{Youngkuk Kim}
\email{youngkuk@skku.edu}
\affiliation{Department of Physics, Sungkyunkwan University, Suwon 16419, Korea}

\author{Moon Jip Park}
\email{moonjippark@hanyang.ac.kr}
\address{Department of Physics, Hanyang University, Seoul 04763, Republic of Korea}
\address{Research Institute for Natural Science and High Pressure, Hanyang University, Seoul, 04763, South Korea}

\date{\today}

\begin{abstract}
Heterostructures of stacked two-dimensional lattices have shown great promise for engineering novel material properties. As an archetypal example of such a system, the hexagon-shared honeycomb-kagome lattice has been experimentally synthesized in various material platforms. In this work, we explore three rotationally symmetric variants of the honeycomb-kagome lattice: the $\textit{hexagonal}$, $\textit{triagonal}$, and $\textit{biaxial}$ phases. While the triagonal and biaxial phases exhibit trivial insulating and Dirac semimetal band structures, respectively, the hexagonal phase hosts a higher-order topological phase driven by band inversion near the $\Gamma$-point. This highlights a key distinction from the conventional band inversions at the $K$-point observed in hexagonal homobilayer systems. Furthermore, we demonstrate how the distinct topological properties of these phases result in network band structures within moiré heterostructures formed by twisted or lattice-mismatched HK systems. These network band structures can be experimentally observed through extrinsic twisting or intrinsic lattice mismatching between the honeycomb and kagome systems.
\end{abstract}

\maketitle

\cyan{Introduction--} A layered heterostructure of two distinct lattice systems provides a versatile platform for the control of electronic properties. When materials with different intrinsic characteristics, such as Fermi levels, orbital symmetries, and lattice structures, are stacked, the contrasting material properties enable modifications to the band structure in a controllable manner\cite{Geim2013, Terrones2013, doi:10.1126/science.aac9439, LI2016322, Castellanos-Gomez2022, Zhang2024}. This hybridization of distinct material properties facilitates novel emergent phenomena in the electronic structure that single-layer or homobilayer structures cannot achieve, enabling new phases of matter driven by tailored interlayer interactions.

\begin{figure}[]
    \centering
    \includegraphics[width=\linewidth]{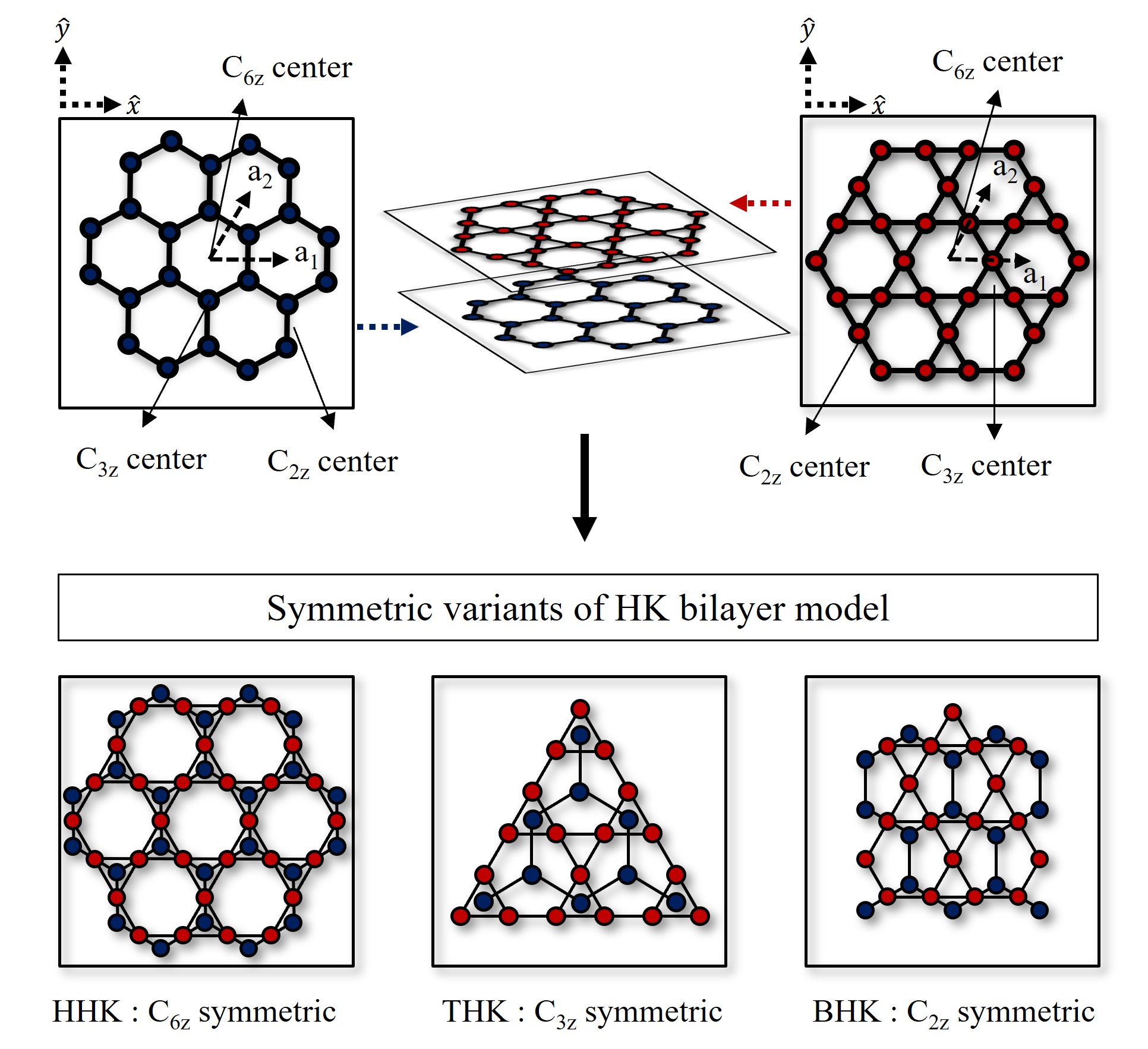}  
    \caption{Honeycomb lattice, kagome lattice and three symmetric variants of HK model. Each structures named HHK, THK, and BHK have \(C_{6z}\), \(C_{3z}\) and \(C_{2z}\) rotational symmetry, respectively.}
    \label{figure1}
\end{figure}

The honeycomb-kagome (HK) bilayer system is a paradigmatic example with numerous material realizations, exhibiting intriguing properties such as Dirac nodal-line semimetals\cite{HK1}, quantum anomalous Hall insulators\cite{HK10,HK3,HK4}, half-metallicity\cite{HK5,HK11}, and valley-induced topological effects\cite{HK15,HK16}.
Many experimental works have demonstrated such HK structures. Examples include HHK-silicene\cite{HK2}, $A_{3}B_{2}$ compounds (A: group-IIB cation, B: group-VA anion)\cite{HK1,HK3,HK4,HK5,HK7,HK13}, superlattice structures such as doped graphene nanomeshes\cite{HK14}, metal-organic frameworks\cite{HK7,HK8,HK9,HK10,HK11,HK12}, and photonic crystal metamaterial platforms\cite{HK6,HK15}. Thus far, these realizations of HK systems predominantly belong to particular stacking structures with C$_{6z}$-rotation symmetry. 

Here, we show that the HK bilayer system supports three high-symmetry stackings, each protected by distinct rotational symmetries (Fig.~\ref{figure1}) with modified interlayer couplings. This symmetry-driven variation enables control over band structures, facilitating the exploration of symmetry-protected topological phases. Additionally, the band hybridization near the band edge, arising from the contrasting Fermi levels of the honeycomb and kagome layers, results in band inversion around the $\Gamma$-point. This is distinct from the conventional Dirac cone physics near the $K$-point in hexagonal homobilayer systems. By controlling symmetry-allowed interlayer couplings, it becomes possible to induce $\Gamma$-point Dirac cones at the Brillouin zone centers, with the resulting band inversion giving rise to higher-order topological insulator (HOTI) phases characterized by protected corner states and non-trivial quadrupole moments.

\begin{figure*}[]
    \centering
    \includegraphics[width=\linewidth]{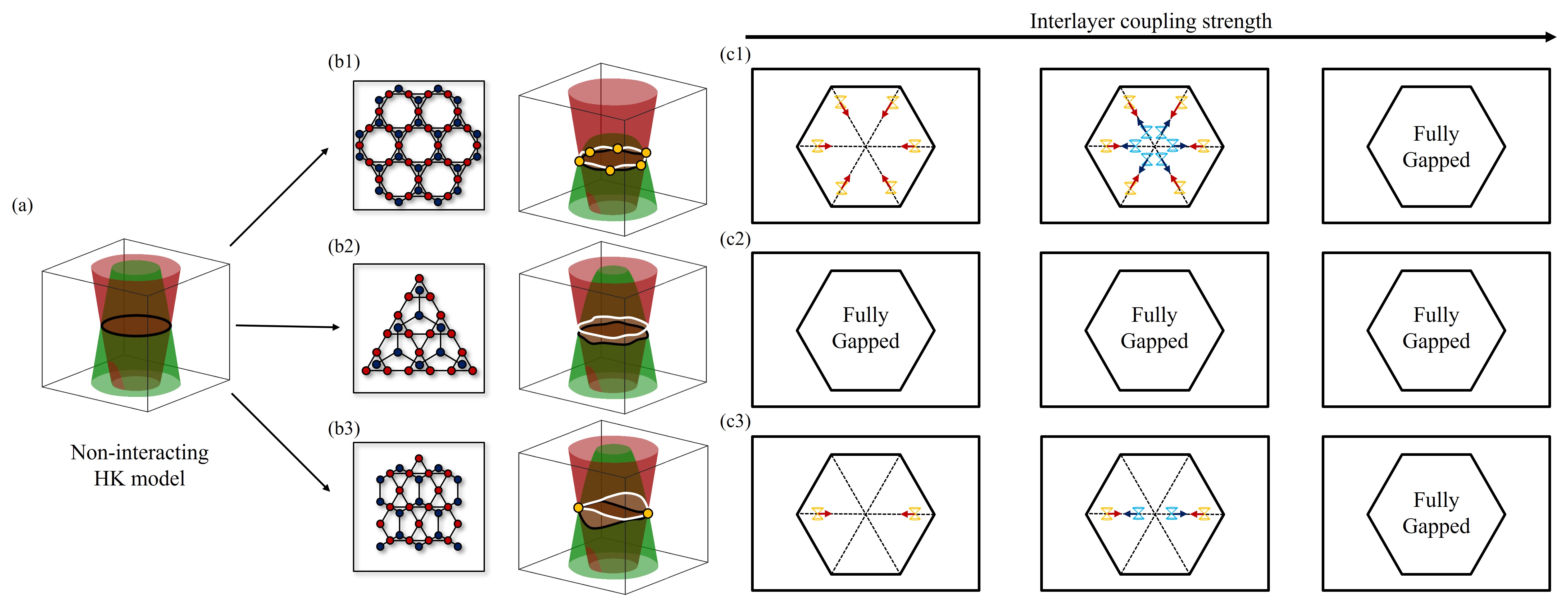}  
    \caption{(a) Dirac nodal line semi-metallic band structure of the non-interacting (or completely prohibited orbital hybridization) case of the HK model. The black line represents the Dirac nodal line. (b1-b3) Three different symmetric variants (HHK, THK, and BHK, respectively) of the HK model and the corresponding hybridized band structure with finite interlayer couplings. The white (black) lines and yellow circles serve as edge lines of the upper (lower) bands and protected Dirac band touching points, respectively. (c1-c3) Schematic figures depicting the dynamics of the Dirac band touching points of HHK, THK, and BHK, respectively, as the interlayer coupling strength increases. Each Dirac touching point (represented by yellow and blue cones) carries a $\pi$-Berry phase and is annihilated by merging with another cone on the symmetric lines (represented by the dashed lines). At sufficiently large interlayer coupling strength, the hybridization of the two bands results in a fully gapped (insulating) phase, despite the preservation of rotational symmetries.
} 
    \label{figure2}
\end{figure*}

\cyan{Symmetric variants of Honeycomb-Kagome lattices--} Starting with the hexagonal HK lattice structure, the other high-symmetry configurations of the HK system can be derived by aligning the rotational centers of each layer. Specifically, two additional symmetry variants of the HK lattice, characterized by $C_{3z}$ and $C_{2z}$ rotational symmetries, can be generated by sliding the kagome lattice relative to the honeycomb lattice, as illustrated in Fig.~\ref{figure1}. Here, the three distinct rotational symmetric variants of the HK lattices are referred to as $hexagonal$-HK (HHK), $triagonal$-HK (THK), and $biaxial$-HK (BHK), respectively, each of which exhibits $C_{6z}$, $C_{3z}$, and $C_{2z}$ rotational symmetries. In addition to the out-of-plane rotational symmetries, the three structures preserves the reflection $\sigma_x$ about in-plane $x$-axis corresponding to the hexagonal, triagonal, orthorhombic space groups (point groups $D_6$, $D_3$, $D_2$), respectively. Based on these crystalline symmetries, the HHK, THK, and BHK lattices manifest the symmetries of the layer groups $p6mm$ (\#77), $p3m1$ (\#69), and $pmm2$ (\#23), respectively.

\cyan{Generic model--} 
The generic form of the symmetry allowed tight-binding model can be written as,
$
\hat{H}=\hat{H}_\textrm{h}+\hat{H}_\textrm{k}+\hat{V}_{\textrm{inter}},
$
where $\hat{H}_{\textrm{h}}$, $\hat{H}_{\textrm{k}}$, and $\hat{V}_{\textrm{inter}}$ denote the Hamiltonians for the intra-layer couplings of the honeycomb and kagome layers, and the inter-layer coupling, respectively. Here, without loss of generality, we use the nearest neighbor intra-layer coupling Hamiltonians, which are expressed as, 
$
\hat{H}_\textrm{h,k} = t_\textrm{h,k}\sum_{<i,j>} (\hat{c}_i^\dagger \hat{c}_j + h.c ) + u_\textrm{h,k}\sum_i \hat{c}_i^\dagger \hat{c}_i$
where $t_\textrm{h,k}$ is the hopping integral and $u_{\textrm{h,k}}$ the onsite potential, with the subscripts h and k indicating the honeycomb and kagome layers, respectively. The notation $\langle , \rangle $ denotes nearest neighbor coupling. The inter-layer coupling is given by the distance dependent hoppings between the sites as,
$
\hat{V}_\textrm{inter} = t_\textrm{hk}\sum_{i,j} e^{-\gamma d_{ij}}(\hat{c}_i^\dagger \hat{c}_j + h.c ) ,
$
where $t_{\textrm{hk}}$ represents the hopping integral for inter-layer coupling. $d_{ij}$ indicates the distance between sites $i$ and $j$, and $\gamma$ characterizes the decaying length of the couplings. \textcolor{red}{--}

\cyan{Electronic properties of hybridization bands--} 
In this work, we consider the honeycomb valence band and the kagome conduction band energetically overlap and hybridization\cite{note1}. The interlayer couplings produces the band hybridization near $\Gamma$-point [See Fig. \ref{figure2} (a)]. In HHK configuration, The line degeneracy of the overlapping bands is gapped out, leaving six Dirac cones with a $\pi$-Berry phase along the six-fold rotationally symmetric mirror-symmetric lines[dahsed lines in Fig. \ref{figure2} (c1)]. Along the mirror symmetric lines, the occurrence of the Dirac cone is insured when $t_h t_k>0$ as the overlapping valence and conduction bands are characterized by different mirror eigenvalues, which prohibit avoided level crossings.

A small sliding of kagome lattice away from C$_{6z}$ symmetric configuration toward THK lattice immediately breaks C$_{2z}$-symmetry gapping out the Dirac cones, realizing the direct band gap insulator in the THK lattice [See Fig. \ref{figure2}(c2)]. On the other hands, in the  BHK configurations, the C$_{2z}$-symmetry is restored, featuring by the pair of Dirac cones on the C$_{2x}$ symmetric lines [See Fig. \ref{figure2}(c3)]. When the orbital hybridization between honeycomb and kagome bands is completely prohibited in the HHK lattice,  the Dirac nodal line semimetal can occur \cite{HK1}. We also observe the nodal-line semimetal phase in the case of a nearest-neighbor interlayer coupling model. However, in general, the inclusion of next-neighbor couplings results in the Dirac nodes.

\begin{figure}[]
    \centering
    \includegraphics[width=\linewidth]{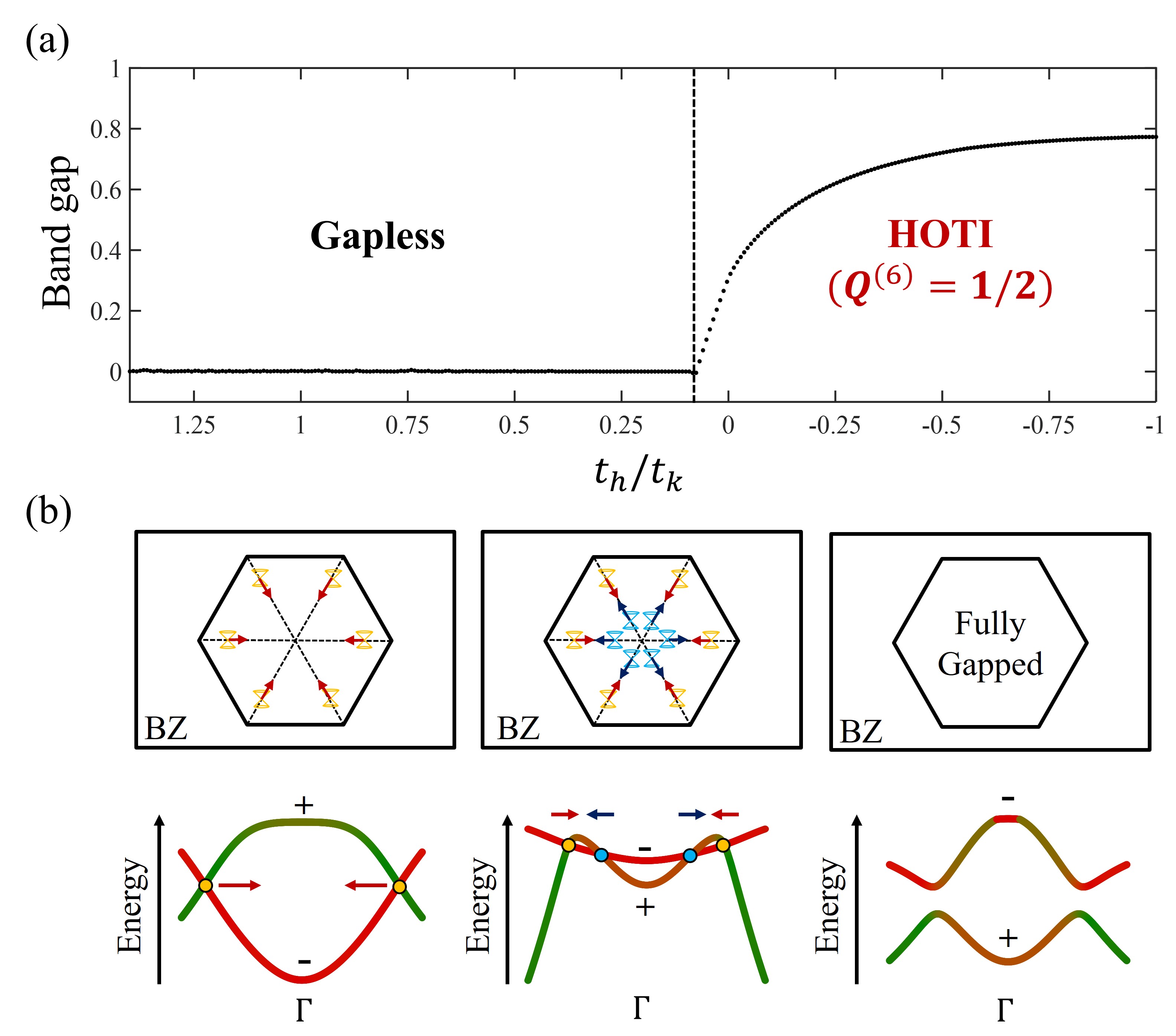}  
    \caption{(a) The phase diagram of the hybridization band gap as a function of the intra-layer coupling ratio ($t_\textrm{h}/t_\textrm{k}$). A higher-order topological insulator (denoted as HOTI) phase is characterized by a quadrupole corner charge of $1/2$. (b) Schematic figures depicting the progression of band hybridization in HHK, as \(t_\textrm{h}\) decreases, corresponding to the phase diagram in (a). The upper panel illustrates the positions and movement of Dirac gaps (yellow and blue cones), and the corresponding band structure along a symmetric line is depicted in the lower panel. The $+$ and $-$ signs represent the mirror eigenvalues of the bands.}
    \label{figure3}
\end{figure}

As the inter-layer coupling strength \( t_\textrm{hk} \) further increases, the Dirac cones in both the HHK and BHK configurations shift along  \(C_{6z}\) (\(C_{2z}\)) symmetric lines toward \(\Gamma\)-point [See Fig.~\ref{figure2}(c)]. Simultaneously, new Dirac cones are generated in pairs at the \(\Gamma\)-point and move outward along the symmetric lines. These newly formed cones eventually merge with the pre-existing Dirac cones, leading to a mutual cancellation of their $\pi$-Berry phases. This merging process opens a global band gap in the band structure. At sufficiently high interlayer coupling, the hybridized bands transition into a fully gapped insulating phase, while the underlying $C_{6z}$ and $C_{2z}$ symmetries.

The insulating phases in the presence of $C_{6z}$ and $C_{2z}$ symmetries accompany the band inversion at $\Gamma$ point. This merging process of the Dirac cones can be also heuristically understood by decreasing the interlayer couplings in the honeycomb lattice \(t_\textrm{h}\). Figure \ref{figure3}(a) shows the topological phase transition of HHK band structure with decreasing  $t_\textrm{h}$. Along the mirror symmetric line, the reduced bandwidth of the honeycomb band produces the additional band crossings (Dirac cones) at $\Gamma$ point.  As the band width approaches to zero, the Dirac cone which moves along the symmetric line, merging with pair-created Dirac cones and eventually becoming gaped out. 

\begin{figure}
    \centering
    \includegraphics[width=\linewidth]{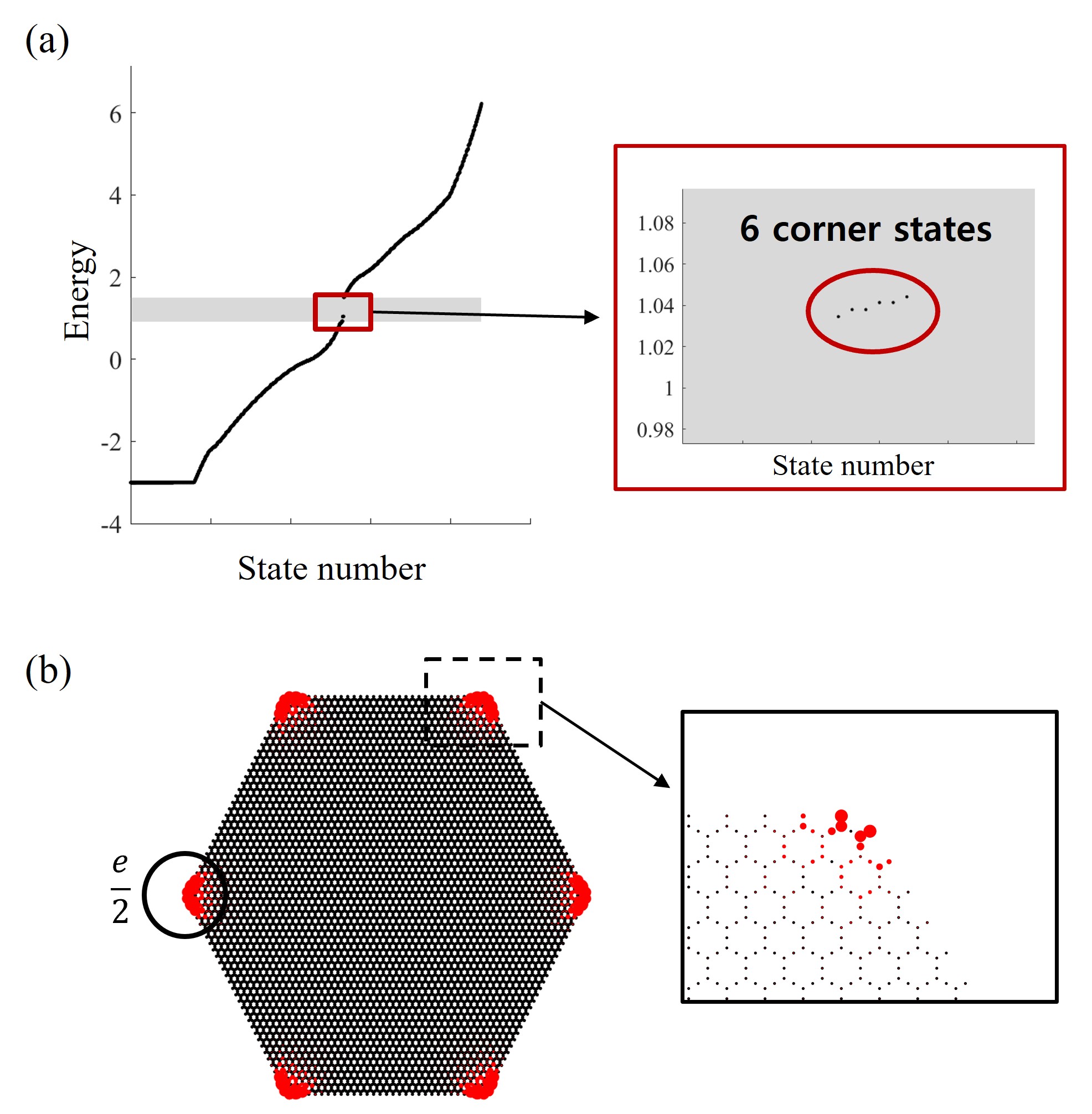}  
    \caption{(a) Energy spectrum for the real-space HHK Hamiltonian under $C_6$-symmetric open boundary conditions, highlighting six corner states within the band gap without edge state. (b) Visualization of the topological corner state, where the size of the red circles represents the amplitudes of the wave functions.}
    \label{figure4}
\end{figure}

\cyan{Higher-order topological insulator in HK model--}
The resulting gapped phase undergoes changes in the mirror eigenvalues, which gives rise to the higher-order topological insulator(HOTI) phase  [See Fig.~\ref{figure3}(b)].
The band gap opening through the merging of $\Gamma$-point Dirac cones exhibits the inversion of the mirror eigenvalues. We find that, while the insulating phase of BHK is topologically trivial, the gapped HHK phase is characterized by the non-trivial quadrupole corner charge as \cite{PhysRevB.99.245151},
\bea
Q^{(6)} = \frac{e}{4} [M_{1}^{(2)}] +\frac{e}{6} [K_{1}^{(3)}] \ \ \textrm{mod} \ e,
\label{C6inv}
\eea
where $[\Pi_{p}^{(n)}]$ represents the difference in the number of occupied bands with $\hat{C}_n$ eigenvalue $e^{2\pi i(p-1)/n}$ between the high-symmetry point $\Pi$ and $\Gamma$. Figure \ref{figure3}(a) and \ref{figure4}(a) shows that the HOTI phase can occur under parameters that allow hybridization of bands within the same mirror symmetry sector. The physical manifestation of the non-trivial quadrupole moment is the topological corner states that appears in the $C_{6z}$-symmetric open boundary condition. Six in-gap corner states are present without any edge state, and the wave function of one of these states is displayed in Fig. \ref{figure4}(b) [See supplementary materials for the calculation of the topological invariant for the THK and BHK lattices].

\begin{figure}
    \centering
    \includegraphics[width=\linewidth]{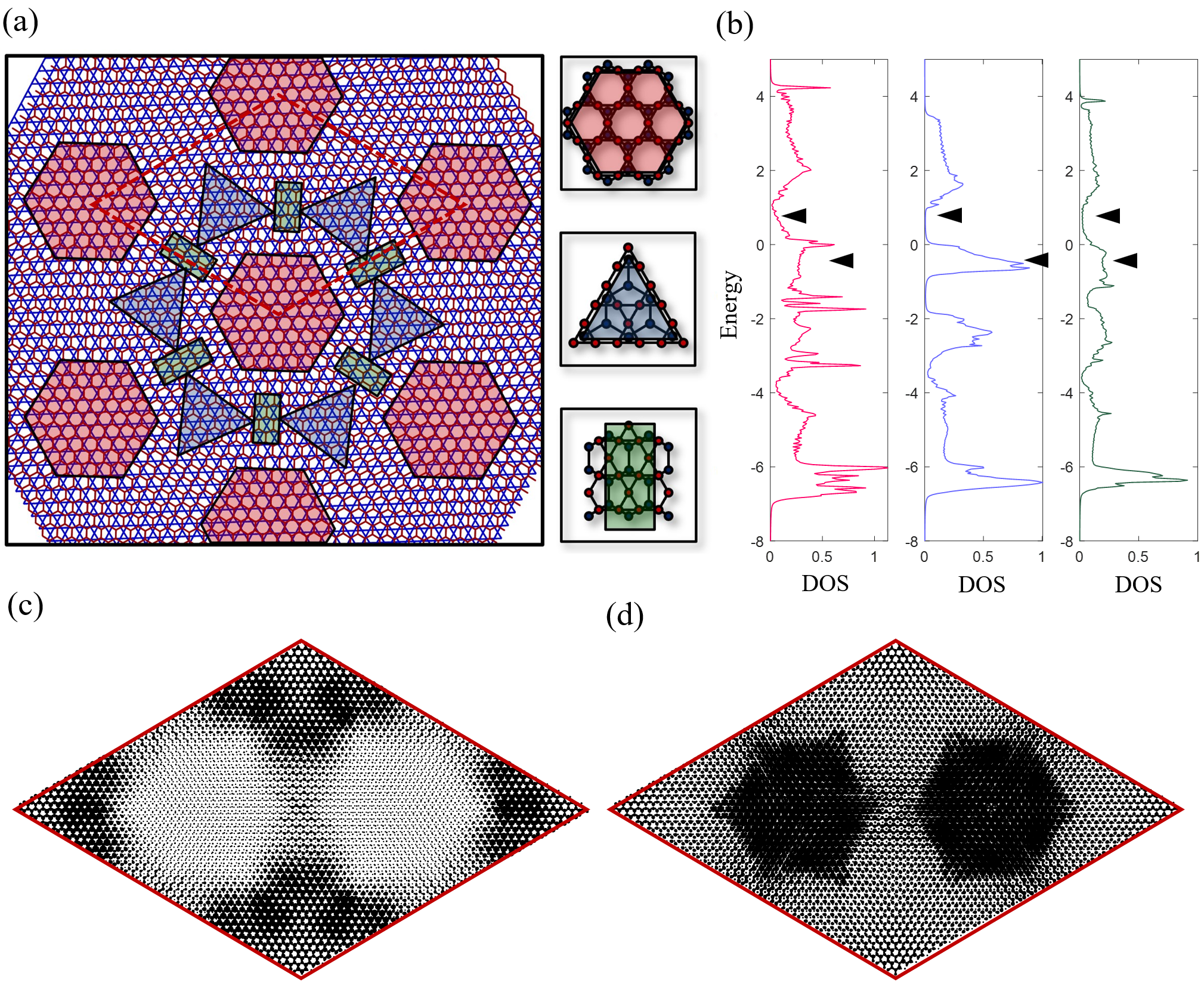}  
    \caption{(a) The moir\'e superlattice structure of TBHK model. The red hexagon, blue triangle and green rectangle region denote the approximate structures of domains-HHK, THK, and BHK respectively. (b) DOS for each local stacking regions in TBHK model. (c) LDOS of TBHK model. The size of the black circles indicate the magnitude of LDOS. 
}
    \label{figure5}
\end{figure}

\cyan{Network model for moiré lattice--} 
The topological metallic and insulating phases identified in the HK model can manifest as experimentally observable electronic states, even when the honeycomb and kagome lattices have slight lattice mismatches. To explore this possibility, we propose the moiré HK model. In the limit of large moiré unit cell, where the primitive moiré translational vector \(\bold{L}_i\) is much larger than the primitive vectors of each layer \(\bold{a}_i\), the moiré HK model can be understood as an arrangement of three symmetric local stacking configurations (HHK, THK, and BHK) as shown in Fig. \ref{figure5}(a).

The commensurate angles $\theta$ of TBHK lattice are given as,
$ \cos{\theta} = {(m^2 + n^2 + 4mn)/2(m^2 + n^2 +mn)}$, where $m$ and $n$ are integers. To obtain the eigenenergy and eigenstate through the diagonalization of the full lattice model, we choose the small angle limit of $\theta = 1.297^\circ$ [$(m,n) = (25,26)$].  The local density of states (LDOS) is calculated as,
\bea
\textrm{LDOS}(\textbf{r},\omega) = -\frac{1}{\pi}\textrm{Im}\left( \sum_{j} \frac{\left| \psi_j(\textbf{r}) \right|^2}{\omega - E_j + i\Gamma} \right),
\label{ldos}
\eea
where $\Gamma$ is an infinitesimal positive number, and the subscript $j$ denotes the eigenstate index. $E_j$ and $\psi_j(\textbf{r})$ are the energy eigenvalue and eigenstate of the Hamiltonian, respectively. 

Fig.~\ref{figure5}(b) shows the DOS for each local stacking region. As anticipated from the untwisted band structure, the THK configuration exhibits a distinct insulating gap near the overlapping energy regions of the honeycomb and kagome bands, while the HHK and BHK configurations display metallic characteristics. These distinct insulating and metallic regions in the moiré structure manifest as spatial variations in the localization of the LDOS.
In the insulating domains corresponding to the THK configurations, the LDOS reveals distinct vacancy regions, indicating suppressed electronic states within the energy gap ($\omega\sim0.8$), as illustrated in Fig.\ref{figure5} (c). Conversely, at the interfaces between these insulating domains, metallic domain walls emerge. The band structure of such a moire structure can be effectively described using a network model. The vacancy in the THK region forms a triangular superlattice with diagonal hoppings. This feature is contrasted from the energy region near kagome conduction bands ($\omega\sim -0.5$), where the enhanced DOS of THK region forms the localized electronic site doublet per unit cell at the THK region. In this case the effective electronic structure is described by the honeycomb superlattice.

\cyan{Discussion--} In the HK bilayer system. the presence or absence of $C_{2z}$ symmetry critically influences the gap opening near $\Gamma$-point. In $C_{2z}$-preserved configurations (HHK, BHK), the insulating phases emerge through Dirac cone merging at strong interlayer coupling regimes. Conversely, broken $C_{2z}$ symmetry, as in the THK configuration, immediately induces a gap under small interlayer interactions. While the similar band hybridization can also occus in the honeycomb-honeycomb or kagome-kagome layers with the onsite difference, we emphasize that these systems exhibit the topologically trivial phases that distinguishes the novel topological properties of HK lattice.

The merging of the Dirac cones at $\Gamma$ point induces the higher-order topological insulating (HOTI) phase in the HHK configuration, featuring non-trivial quadrupole corner charges. This property extends the scope of HK bilayer systems to include materials such as $\text{Mg}_3\text{C}_2$ and $\text{As}_2\text{O}_3$, and various organic frameworks. However, other factors such as spin-orbit coupling, and structural deformations are crucial for accurate topological predictions. Notably, our search of 2D materials databases\cite{petralanda2024,jiang2024} revealed that the HHK configuration is realized in $\text{As}_2\text{O}_3$ [See supplementary materials for DFT calculation]. While the electron filling in this material differs significantly from the conditions required to access a topological gap, the discovery of such an atomic structure reinforces the prospect of finding HK lattice materials in future studies.

Furthermore, we explored the moiré structure of the HK lattice and demonstrated that the electronic properties of the local regions in the moiré HK model can be estimated based on the symmetric stacking configurations of HK lattices. For a sufficiently large unit-cell moiré HK model, we can exploit the superlattice electronic structure arising from the distinct electronic properties of HK lattices induced by structural variation. One example is the formation of robust metallic domain walls that separate different regions. In general, such characteristics appear not only in the twisted HK model but also in general moiré structures, including those formed due to intrinsic lattice mismatch. We also investigate the moiré structure induced by a slight discrepancy in the lattice constant. Similar to the TBHK case, we confirm that the electronic properties of local regions in the moiré HK model can be estimated based on the stacking configurations of HK lattices. Additionally, we observe the emergence of domain wall networks that separate two distinct local regions [See supplementary materials]. Our work lays the groundwork for realizing heterostructures with mesoscopic network domains and advancing the discovery of new topological materials in honeycomb kagome heterostructures, as well as understanding electronic structures arising from various structural origins.

\bibliography{reference}

\section{Acknowledgement} 
This work was supported by the National Research Foundation of Korea
(NRF) grant funded by the Korea government (MSIT) (Grants No. RS-2023-00218998). This work was supported by the BK21 FOUR (Fostering Outstanding Universities for Research) program through the National Research Foundation (NRF) funded by the Ministry of Education of Korea.

\pagebreak

\newpage

\renewcommand{\thefigure}{S\arabic{figure}}
\setcounter{figure}{0}
\renewcommand{\theequation}{S\arabic{equation}}
\setcounter{equation}{0}

\begin{widetext}

\section{Supplemental Material for \\ "Stacking-dependent Topological Electronic Structures in Honeycomb-Kagome Heterolayers"}

\maketitle

\section{Stacking dependent electronic structure for the other case}
In the main article, we consider the hybridization between two bands: the conduction band of the kagome lattice and the valence band of the honeycomb lattice. The hybridization energy level is modulated by the onsite energy terms in each lattice's Hamiltonian. Another case of interest is the hybridization between the honeycomb conduction band and the kagome valence band, which includes the quadratic touching flat band [see Fig. \ref{supple1}(a)]. The change in electronic properties in this case can be interpreted using the same Dirac cone dynamics as in the main article. However, the presence of the flat band renders the hybridization process more intricate.

Similar to the case discussed in the main article, the electronic properties of symmetric variants are significantly influenced by the presence of $C_{2z}$ symmetry, which protects the Dirac cones at rotationally symmetric points. Unlike the HHK and BHK cases, the THK lattice immediately breaks $C_{2z}$ symmetry, leading to the gapping out of the Dirac cones and realizing a direct band gap insulator in the THK lattice [see Fig. \ref{supple1}(b2)]. On the other hand, in the HHK and BHK configurations, $C_{2z}$ symmetry protects the Dirac cones, resulting in a Dirac semimetallic band structure.

However, as the interlayer coupling strength increases, the presence of the flat band and the quadratic band-touching point induces deviations from the case discussed in the main article. In the HHK and BHK configurations, the Dirac cones between the kagome valence band and the honeycomb conduction band shift toward the $\Gamma$-point along the symmetric line, eventually transitioning from the kagome valence band to the kagome flat band.

The quadratic band-touching point is protected by time-resvesal symmetry and rotational symmetry\cite{PhysRevLett.103.046811}, unlike in the case discussed in the main article, where the semimetallic behavior of the THK lattice is restored in an intermediate interlayer coupling strength regime [see Fig. \ref{supple1}(b2)].

However, at sufficiently large interlayer coupling strength, the Dirac cones between the flat band and the honeycomb conduction band move along the high-symmetry line and eventually merge. Consequently, as in the main article, the hybridization of the bands leads to the formation of a fully gapped (insulating) phase, despite the preservation of rotational symmetries.

\begin{figure*}[ht]
    \centering
    \includegraphics[width=\linewidth]{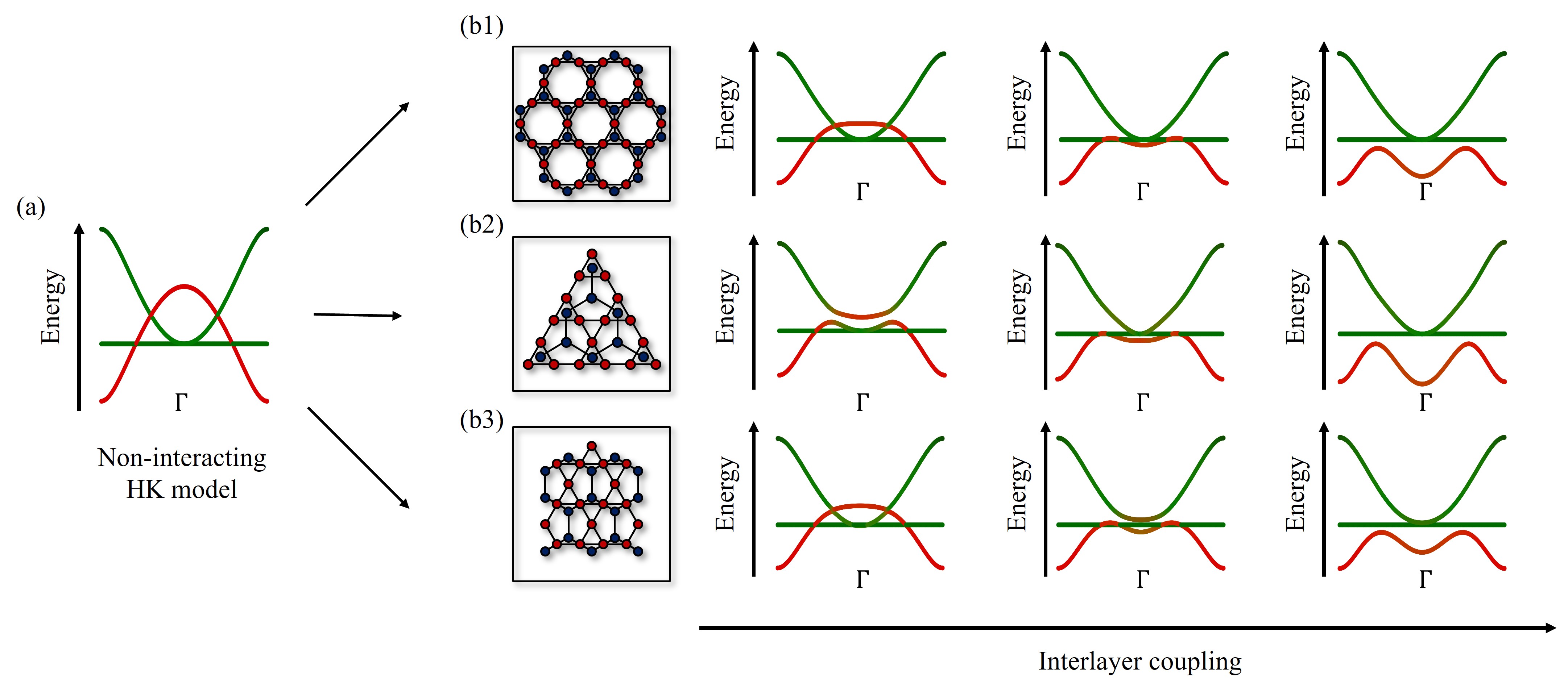}  
    \caption{(a) Dirac nodal line semimetallic band structure in the non-interacting limit or in the absence of orbital hybridization in the HK model. The one-dimensional band structure shown in the figure is a cross-section of the two-dimensional band structure along the symmetry line. (b1–b3) Three distinct symmetric variants of the HK model (HHK, THK, and BHK, respectively) and schematic representations of the corresponding hybridized band structures along a symmetric line as the interlayer coupling strength increases. At sufficiently large interlayer coupling strength, the hybridization of the two bands leads to the emergence of a fully gapped (insulating) phase, even in the presence of rotational symmetries.
    }
    \label{supple1}
\end{figure*}

\section{Calculation of topological invariants}
For time-reversal and rotationally symmetric lattice systems in class AI of the ten-fold classification, W. A. Benalcazar, et al. have suggested efficient way to calculate quantized dipoles moment and quadrupole moments using the eigenvalues of rotational operators in momentum space.\cite{PhysRevB.99.245151} In the case of a system with six-fold rotational symmetry(HHK), the quantized dipole moment $\bold{P}^{(6)}$is always trivial, and the quadrupole corner charge can be explicitly calculated using the formula as
\bea
Q^{(6)} = \frac{e}{4} [M_{1}^{(2)}] +\frac{e}{6} [K_{1}^{(3)}] \ \textrm{mod} \ e,
\label{C6inv}
\eea

where $[\Pi_{p}^{(n)}]$ represents the difference in the number of occupied bands with $\hat{C}_n$ eigenvalue $e^{2\pi i(p-1)/n}$ between the high-symmetry point $\Pi$ and $\Gamma$. Similarly with $C_6$ case, we can also calculate dipole moment invariant $\bold{P}^{(n)}$ and quadrupole charge invariant $Q^{(n)}$ for $n=3$ and $n=2$ symmetric cases with the formulae
\begin{alignat}{2}
    \bold{P}^{(2)} &= {e\over2}\left(\left[Y_1^{(2)}\right] +  \left[M_1^{(2)}\right] \right) \bold{a}_1 \nonumber\\ &\ \ + {e\over2}\left(\left[X_1^{(2)}\right] +  \left[M_1^{(2)}\right] \right) \bold{a}_2\ \textrm{mod} \ e \nonumber  \\
    \bold{P}^{(3)} &= {2e\over3}\left(\left[K_1^{(3)}\right] + 2 \left[K_2^{(3)}\right] \right) ( \bold{a}_1 + \bold{a}_2)\ \textrm{mod} \ e \nonumber\\
    Q^{(2)} &= {e\over4}\left(-\left[X_1^{(2)}\right] -\left[Y_1^{(2)}\right]+ \left[M_1^{(2)}\right] \right)\ \textrm{mod} \ e \nonumber\\
    Q^{(3)} &= {e\over3}\left(\left[K_2^{(3)}\right]  \right)\ \textrm{mod} \ e. \label{eq:last}
\end{alignat}

Topological invariants $\bold{P}^{(n)}$ and ${Q}^{(n)}$ of THK and BHK both cases are trivial regardless of the parameters used in this work. Note this invariants depend to lattice geometries(rotation operator in lattice) and the hierarchical order of bands(for calculating $[\Pi_{p}^{(n)}]$). Therefore, if the lattice structure is well defined and the change of parameter does not reverse the order of the bands, then the topological invariants will not be varied.

\section{Moir\'e structure induced by lattice constant discrepancy}
In the main text, we demonstrated a local electronic structures and domian wall network state in the TBHK model. However, fabricating layered honeycomb and kagome two-dimensional materials with identical lattice constants poses significant experimental challenges. A discrepancy between the primitive vectors of the layers produces a moiré pattern. Here, we focus on the case in which the kagome lattice unit cell is slightly larger than the honeycomb lattice unit cell, and we call this model as "lattice mismatched HK model".

We define the primitive vector of the kagome lattice $\bold{a}_i^{(k)}$ in terms of the primitive vector of the honeycomb lattice $\bold{a}_i$ and a certain integer $N$ as $\bold{a}_i^{(k)} = \left(1 + 1/N \right)\bold{a}_i$. Consequently, the primitive moiré vectors are given by $\bold{L}_i = (1+N)\bold{a}_i$. For our numerical calculations, we choose $N=50$, corresponding to the large moiré unit cell limit with $\left| \bold{L}_i \right|/\left| \bold{a}_i \right| = 51$ and encompassing more than 10,000 sites. In this regime, the moiré structure can be effectively treated using the continuum approximation of the network model, analogous to the TBHK case discussed in the main text. Figure \ref{supple2}(a) illustrates the moiré pattern and the distinct local stacking configurations—HHK, THK, and BHK.

Figure \ref{supple2}(b) presents the density of states (DOS) calculated for each local stacking region, which are highlighted by the red hexagon, blue triangle, and green rectangle in Fig. \ref{supple2}(a). Consistent with the TBHK model, the THK configuration exhibits a clear insulating gap near the energy overlap of the honeycomb and kagome bands, whereas the HHK and BHK configurations remain metallic. The corresponding energy level($\omega \sim 0.5$) is indicated by the triangle arrow in Fig. \ref{supple2}(b). Furthermore, these distinct insulating and metallic domains manifest in the spatial distribution of the local density of states (LDOS). In the THK regions, the LDOS displays pronounced vacancy areas, indicating suppressed electronic states within the gap at ($\omega \sim 0.5$)[See Fig. \ref{supple2}(c)].

Overall, the LDOS calculations indicate that the electronic properties in the lattice-mismatched case are similar to those in the TBHK model, suggesting that our findings can be generalized to other large-unit-cell moiré structures constructed from honeycomb and kagome lattices.

\begin{figure*}[ht]
    \centering
    \includegraphics[width=\linewidth]{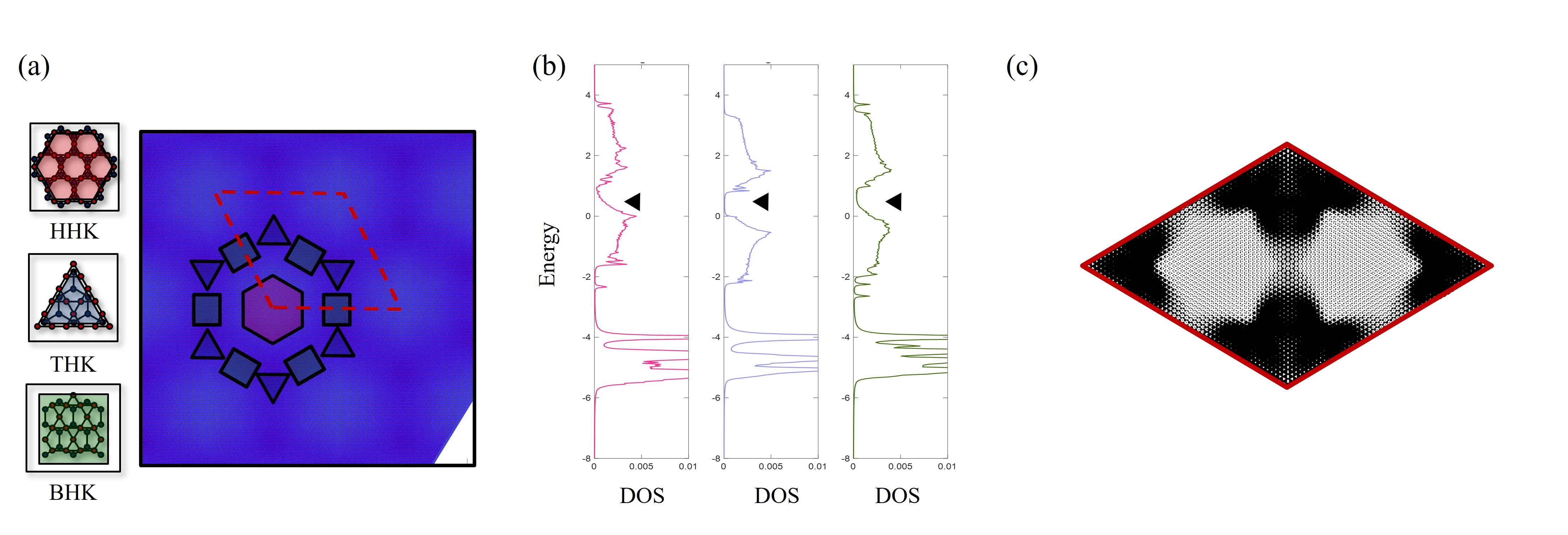}  
    \caption{(a) The moir\'e superlattice structure of the Lattice mismatch case. The red hexagon, blue triangle, and green rectangle region denote the approximate structures of domains-HHK, THK, and BHK respectively. (b) DOS for each local stacking regions in lattice mismatched moir\'e-HK model. (c) LDOS of lattice mismatched moir\'e-HK model. The size of the black circles indicate the magnitude of LDOS.  
    }
    \label{supple2}
\end{figure*}

\section{Material realization for energetically overlap HHK model}

\subsection{Material Realization}

Motivated by the theoretical exploration of the HK model, we have searched for the materials that realize the honeycomb-kagom\'e lattices. We notice that the HK lattices exhibit rotational symmetry along the out-of-plane $z$-direction, in addition to the mirror reflection symmetry along the in-plane $x$-direction. Based on these crystalline symmetries, we determine that the HHK, THK, and BHK lattices manifest the symmetries of the layer groups $p6mm$ (\#77), $p3m1$ (\#69), and $pmm2$ (\#23), respectively. By examining these layer groups within the Topological 2D Materials Database \cite{petralanda2024, jiang2024}, we identify As$_2$O$_3$ as a candidate for the HHK structure, while no suitable options are found for the THK or BHK lattices. 

The unit cell of As$_2$O$_3$ is illustrated in Figure 1 (a). The structure consists of two formula units of arsenic atoms forming a honeycomb layer, while three formula units of oxygen atoms align to create a Kagome layer. The Separation between the honeycomb and Kagome layers is 0.94 \AA. Fig.1 (b) and (c) present the electronic structure of As$_2$O$_3$, obtained from the first-principles density functional theory (DFT) calculation. The occupied bands near the Fermi level in Fig.1 (b) exhibit features of hybridized Kagome bands, characterized by the flattened topmost occupied bands. In contrast, Fig.1 (c) shows distinct Kagome bands at lower energy, likely originating from the low-energy states of oxygen orbitals in the Kagome layer.

We calculated the orbital-projected band structure to further investigate the hybridization of honeycomb and Kagome bands near the Fermi level. Figure 2 illustrates the fat band plot, where the thickness of lines scales with the orbital projections. Our results indicate that the $p_z$-orbitals of oxygen and arsenic atoms contribute to the five valence bands near the Fermi level, supporting the mixing of electronic states originating from distinct layers. Additionally, the projected band structure reveals that the low-energy Kagome bands in Fig.2 (f)-(h) predominantly originate from the $s$-orbitals of oxygen atoms.

The electronic structure of As$_2$O$_3$ suggests the appealing idea of modifying the chemical composition of the compound to observe hybridization of HK bands in real materials. Previous experiments have demonstrated that such manipulation is feasible via the doping of nitrogen or oxygen atoms onto a Graphene sheet \cite{lu2022tunable}. We also propose that inducing oxygen vacancies in a triangular lattice could serve as a promising approach to realize the HK structure in synthesizable materials \cite{wang2024high}.

\begin{figure}
    \centering
    \includegraphics[width=0.8\textwidth]{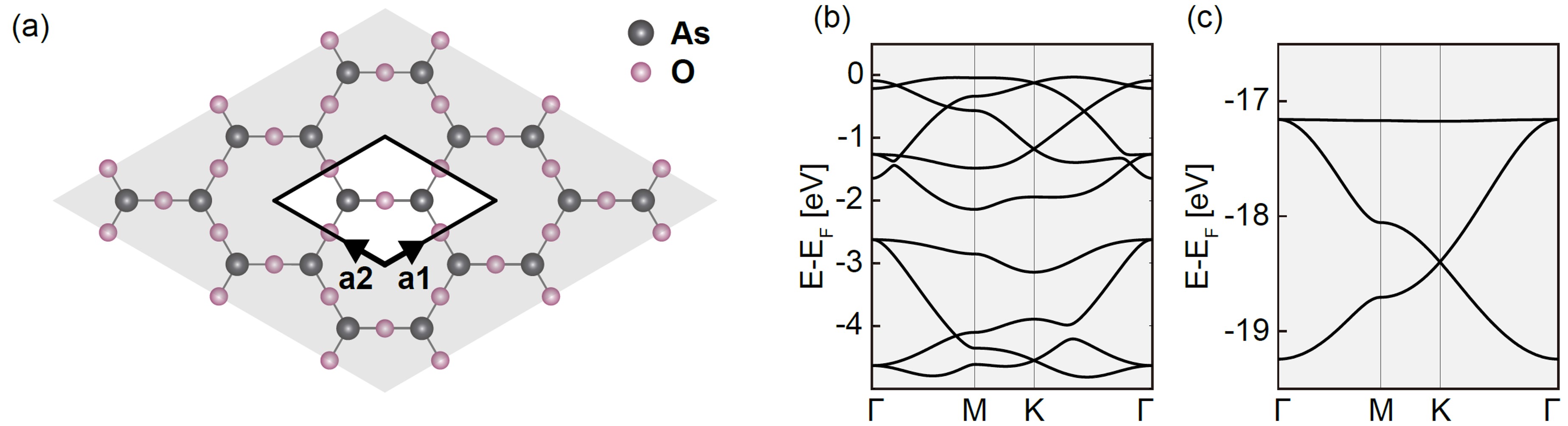}
\caption{\label{fig:mat}
The atomic and electronic structure of candidate material As$_2$O$_3$. (a) Top view of the atomic structure. The gray (purple) solid circles represent arsenic (oxygen) atoms. The black solid line in the middle of the figure indicates the single unit cell. (b) Occupied electronic bands near the Fermi level. (c) Kagome bands in the lower energy region.
}\end{figure}

\begin{figure}
    \centering
    \includegraphics[width=1.0\textwidth]{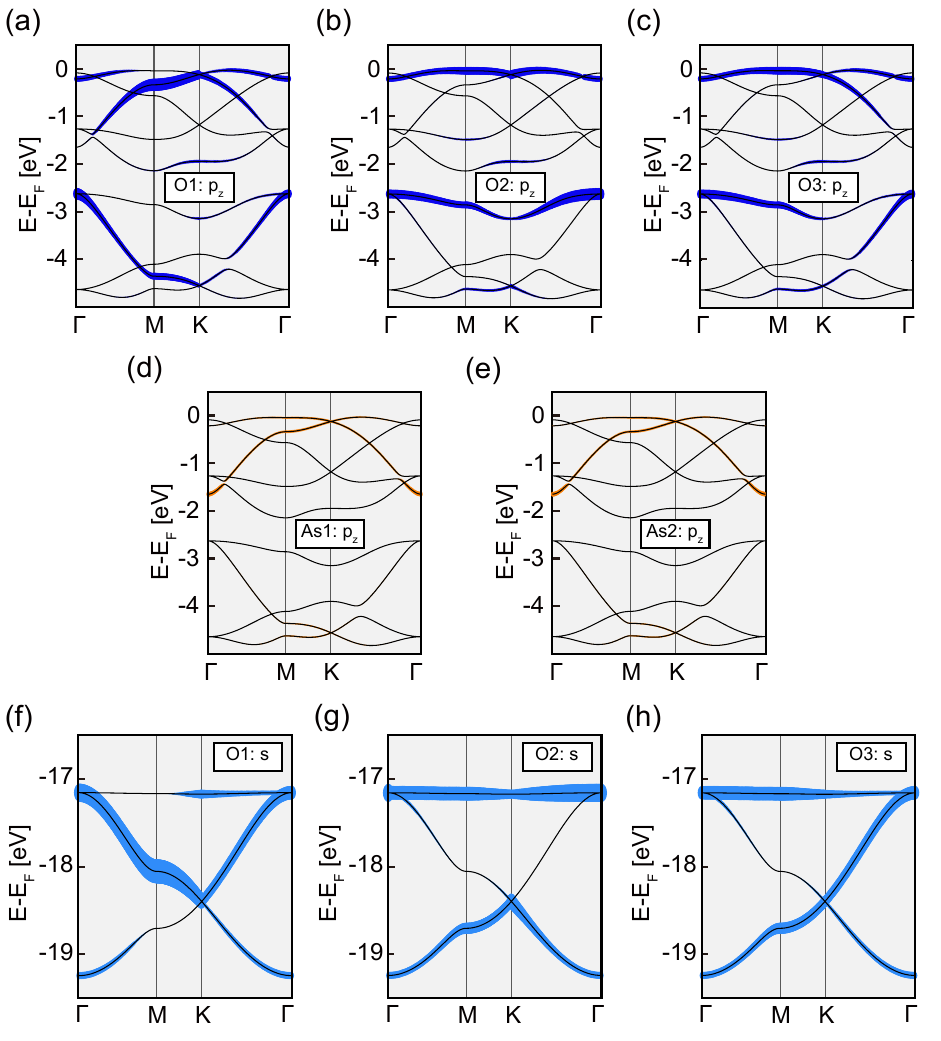}
\caption{\label{fig:pdos}
Fat band plot of the orbital-projected band structure. The line thickness represents the contribution of projected orbitals. (a)–(c) Blue fat bands near the Fermi level show the $p_z$-orbital projections of oxygen atoms in the Kagome layer. (d), (e) Orange fat bands correspond to the $p_z$-orbital projections of arsenic atoms. (f)-(h) Light blue fat bands illustrate the $s$-orbital contributions to the low-energy Kagome bands from oxygen atoms.
}\end{figure}

\subsection{Computational Method}

Our first-principles calculations based on the density functional theory were performed using the \textsc{vasp} package \cite{PhysRevB.54.11169}. The atomic Structure of As$_2$O$_3$ was obtained from the Topological 2D Materials Database \cite{petralanda2024, jiang2024}. The lattice constants were set to $a = 5.43$ \AA, with the separation between the honeycomb and Kagome layers fixed at $0.94$ \AA. To model the monolayer system, a vacuum distance of $13.02$ \AA ~ was introduced along the out-of-plane direction. The self-consistent charge density was obtained on a $16 \times 16 \times 1$ Monkhorst-Pack momentum grid \cite{monkhorst1976special}, with an energy cutoff of $400$ eV for the plane-wave basis set. The total energy was successfully converged below $10^{-7}$ eV criterion.

\end{widetext}

\end{document}